\documentclass[12pt]{article}
\usepackage{epsfig}
\begin{document}
\begin{center}
{\Large {\bf The observability of an additional discrete symmetry in the
Standard Model}

\vskip-40mm \rightline{\small ITEP-LAT/2006-08} \vskip 30mm

{
\vspace{1cm}
{ M.A.~Zubkov$^a$ }\\
\vspace{.5cm} {\it $^a$ ITEP, B.Cheremushkinskaya 25, Moscow, 117259, Russia
}}}
\end{center}

\begin{abstract}
Standard Model may be defined with the additional discrete symmetry, i.e. with
the gauge group $SU(3)\times SU(2) \times U(1)/{\cal Z}$ (${\cal Z} = Z_6$,
$Z_3$ or $Z_2$) instead of the usual $SU(3)\times SU(2) \times U(1)$. It has
the same perturbation expansion as the conventional one. However, it may
describe nature in a different way at the energies compared to the triviality
bound (of about $1$ Tev). In this paper we present a possibility to observe
this difference assuming that the gauge group of the Standard Model is embedded
into the gauge group of an {\it a priory} unknown model, which describes
physics at a Tev scale. This difference is related to the monopole content of
the theory. We illustrate our results by consideration of the Petite
Unification of quarks and leptons.
\end{abstract}


New physics is coming at a Tev scale. This becomes evident due to the existence
of the triviality bound on the validity of the Standard Model
\cite{M_H,TEV,Extention}. Namely, the Standard Model (SM) clearly does not work
at the energies above about $1$ Tev. Therefore, at this scale some other theory
should appear, which incorporates Standard Model as a low energy approximation.

Long time ago it was recognized that the spontaneous breakdown of $SU(5)$
symmetry in the Unified model actually leads to the gauge group $SU(3)\times
SU(2) \times U(1)/Z_6$ instead of the conventional $SU(3)\times SU(2) \times
U(1)$ (see, for example, \cite{Z6} and references therein). The appearance of
the additional $Z_6$ symmetry in the fermion and Higgs sectors of the Standard
Model itself was recovered later within the lattice field theory
\cite{BVZ2003,BVZ2004, BVZ2005, BVZ2006}. Independently $Z_6$ symmetry in the
Higgs sector of the Standard Model was considered in \cite{Z6f}.

Thus we are faced the following question. What is the gauge group of the
Standard Model? It may be either $SU(3)\times SU(2) \times U(1)$ or
$SU(3)\times SU(2) \times U(1)/{\cal Z}$, where ${\cal Z}$ is equal to $Z_6$,
or to one of its subgroups: $Z_3$ or $Z_2$. But first we must understand if
there is any difference between the models or not.

On the level of perturbation expansion those versions of the Standard Model are
identical. In \cite{BVZ2003} the supposition was made that actually those
models may differ due to the nonperturbative effects. The lattice simulations
show that there is indeed some difference in the lattice realizations of the
Standard Model with the gauge groups $SU(3)\times SU(2) \times U(1)/Z_6$ and
$SU(3)\times SU(2) \times U(1)$ both at zero and finite temperature. However,
there is no evidence that those differences survive in the continuum theory.

In this letter we take into account that the Standard Model describes nature
only up to the energies of about a few  Tev. This can have an effect on the
topology of the Standard Model. Namely, there may appear small regions of sizes
of the order of $1 \, {\rm Tev}^{-1}$, where the conventional fields of the
Standard Model are not defined. These regions may represent point-like or
string - like objects\footnote{Actually these objects are the soliton-like
states of the theory, which describes Tev scale physics.}.  As a result we must
consider the topology of the Standard Model
 within the space-time manyfold $\cal M$ with nontrivial $\pi_2({\cal M})$
and (or) $\pi_1({\cal M})$. We shall explain here, that the mentioned objects
may appear with masses of about $40$ Tev.

There are several patterns of unification of interactions, which were
considered up to now. Among them we should mention at least two examples, in
which gauge group of the Standard Model is extended already at the Tev scale.
Namely, in the so - called Little Higgs models \cite{little_higgs} $SU(2)\times
U(1)$ subgroup is embedded into a larger group, which is gauged partially. The
correspondent symmetry is broken at a few Tev. Then some of the Nambu -
Goldstone bosons become massive due to the radiative corrections and play the
role of the Higgs field of the Standard Model. The topological objects, which
appear in the Little Higgs model were considered in \cite{TopLittleHiggs} for
the case of the so-called Littlest Higgs model.

The second example is the so-called Petite Unification (see, for example,
\cite{PUT,PUT1} and references therein). In the correspondent models the gauge
symmetry of the Standard Model is extended to a larger one at the Tev scale.
The resulting models have two different coupling constants correspondent to
strong and Electroweak interactions unlike Grand Unified models, in which there
is only one coupling constant and the unification is achieved at the GUT scale
$10^{15}$ Gev.

It is not important for us, which particular model describes Tev scale physics.
The only important feature of such a model is that the gauge group of the
Standard Model is embedded into the gauge group of the unified model and the
latter is large enough. It will be clear later to what extent it should be
larger than $SU(3)\times SU(2) \times U(1)/{\cal Z}$.

Let us now fix the closed surface $\Sigma$ in $4$-dimensional space $R^4$. For
any closed loop $\cal C$, which winds around this surface, we may calculate the
Wilson loops $\Gamma = {\rm P} \,  {\rm exp} (i\int_{\cal C} C^{\mu}
dx^{\mu})$, $U = {\rm P} \, {\rm exp} (i\int_{\cal C} A^{\mu} dx^{\mu})$, and
$e^{i\theta} =  {\rm exp} (i\int_{\cal C} B^{\mu} dx^{\mu})$, where $C$, $A$,
and $B$ are correspondingly $SU(3)$, $SU(2)$ and $U(1)$ gauge fields of the
Standard Model. In the usual realization of the Standard Model with the gauge
group $SU(3)\times SU(2) \times U(1)$ such Wilson loops should tend to unity,
when the length of $\cal C$ tends to zero ($|{\cal C}| \rightarrow 0$).
However, in the $SU(3)\times SU(2) \times U(1)/{\cal Z}$ gauge theory the
following values of the Wilson loops are allowed at $|{\cal C}| \rightarrow 0$:
\begin{eqnarray}
\Gamma &=& {\rm P} \, {\rm exp} (i\int_{\cal C} C^{\mu} dx^{\mu}) = e^{N
\frac{2\pi
i}{3}}\nonumber\\
U &=& {\rm P} \, {\rm exp} (i\int_{\cal C} A^{\mu} dx^{\mu}) =
e^{-N \pi i}\nonumber\\
e^{i\theta} &=& {\rm exp} (i\int_{\cal C} B^{\mu} dx^{\mu}) = e^{N \pi
i},\label{Sing}
\end{eqnarray}
where $N = 0,1,2,3,4,5$ for ${\cal Z}=Z_6$, $N = 0,2,4$ for ${\cal Z}=Z_3$, and
$N = 0,3$ for ${\cal Z}=Z_2$. Then the surface $\Sigma$ may carry $Z_2$ flux
$\pi [N\, {\rm mod}\,2]$ for ${\cal Z} = Z_2, Z_6$. It also may carry $Z_3$
flux $\frac{2\pi [N\, {\rm mod}\,3]}{3}$ for ${\cal Z} = Z_3, Z_6$.

Any configuration with the singularity of the type (\ref{Sing}) could be
eliminated via a singular gauge transformation. Therefore the appearance of
such configurations in the theory with the gauge group $SU(3)\times SU(2)
\times U(1)/{\cal Z}$ does not influence the dynamics.

Now let us consider an open surface $\Sigma$. Let the small vicinity of its
boundary $U(\partial \Sigma)$ represent a point - like soliton state of the
unified theory. This means that the fields of the Standard Model are defined
now everywhere except $U(\partial \Sigma)$. Let us consider such a
configuration, that for infinitely small contours $\cal C$ (winding around
$\Sigma$) the mentioned above Wilson loops are expressed as in (\ref{Sing}).
For $N \ne 0$ it is not possible to expand the definition of such a
configuration to $U(\partial \Sigma)$. However, this could become possible
within the unified model if the gauge group of the Standard Model $SU(3)\times
SU(2) \times U(1)/{\cal Z}$ is embedded into the simply connected group $\cal
H$. This follows immediately from the fact that any closed loop in such $\cal
H$ can be deformed smoothly to a point and this point could be moved to unity.
Actually, for such $\cal H$ we have $\pi_2({\cal H}/[SU(3)\times SU(2) \times
U(1)/{\cal Z}]) = \pi_1(SU(3)\times SU(2) \times U(1)/{\cal Z})$. This means
that in such unified model the monopole-like soliton states are allowed. The
configurations with (\ref{Sing}) and $N\ne 0$ represent fundamental monopoles
of the unified model\footnote{Actually these configurations were already
considered (see, for example \cite{Z6}, where they represent fundamental
monopoles of the $SU(5)$ unified model). However, in \cite{Z6} it was implied
that such soliton states could appear with the masses of the order of GUT scale
($10^{15}$ Gev). In our case the appearance of such objects is expected already
at the energies compared to $1$ Tev because we consider the unified model, in
which $\cal H$ is broken to the gauge group of the Standard Model already at
this scale.}. The other monopoles could be constructed of the fundamental
monopoles as of building blocks.  In the unified model, which breaks down to
the SM with the gauge group $SU(3)\times SU(2) \times U(1)$ such configurations
for $N\ne 0$ are simply not allowed. This gives us the way to distinguish
between the two versions of the Standard Model.

The unified model, which breaks down to the SM with the gauge group
$SU(3)\times SU(2) \times U(1)$ also contains monopoles because $\pi_2({\cal
H}/[SU(3)\times SU(2) \times U(1)]) = \pi_1(SU(3)\times SU(2) \times U(1)) =
Z$. They correspond to the Dirac strings with $ \int_{\cal C} B^{\mu} dx^{\mu}
= 6 \pi K, K\in Z$ and should be distinguished from the monopoles  of the SM
with the additional discrete symmetry via counting their hypercharge $U(1)$
magnetic flux.

We should mention here that another monopoles were shown to exist in the
Standard Model. These are the so - called Nambu monopoles (see, for example,
\cite{Nambu} and references therein). There are several aspects, in which they
are different from the objects considered in this paper. First, they are not
topologically stable. Therefore they may appear only in the combination of
monopole - antimonopole pair, which is connected via the so-called Z - string.
Next, they have fractional electromagnetic charge. Electromagnetic field is
expressed through $A$ and $B$ as follows:
\begin{equation}
 A_{\rm em}  =  2 B - 2 \,{\rm sin}^2\, \theta_W (A_3+B).
\label{A_em}
\end{equation}
The net hypercharge magnetic flux of Nambu monopoles is zero. Therefore the
electromagnetic flux is proportional to $4 \pi \,{\rm sin}^2\, \theta_W$. The
monopoles, which were considered above have nontrivial hypercharge flux and
have electromagnetic flux proportional to $2\pi$.

Another type of monopole, which was considered within the Standard Model is the
Cho-Maison monopole\cite{Cho}. Monopoles of this kind were shown to have
hypercharge flux $2\pi$ and electromagnetic flux $4\pi$. They have infinite
self energy assuming that the Standard Model has an infinite cutoff. All that
allow us to identify them with the monopoles of the unified model, which breaks
down to the SM with the gauge group  $SU(3)\times SU(2) \times U(1)$. They may
exist due to the finiteness of the ultraviolet cutoff of the SM.

Using an analogy with t'Hooft - Polyakov monopoles\cite{HooftPolyakov} we can
estimate masses of the fundamental monopoles, which have nontrivial $U(1)$
flux, to be of the order of $\frac{e \Lambda N}{\alpha} \sim 40 N$ Tev, where
$\Lambda \sim 1 \, {\rm Tev}$ is the scale of the breakdown and $\alpha =
\frac{e^2}{4\pi}$
 is the fine structure constant ($\alpha(M_Z)\sim \frac{1}{128}$) (see, for example, \cite{Weinb},
 where monopoles were considered for an arbitrary compact simple gauge group in the BPS limit).
According to (\ref{A_em}) usual magnetic flux of the fundamental monopoles is
equal to $2\pi N$. If  Standard Model has the conventional gauge group then
monopoles will appear with $N$ proportional to $6$, and the magnetic flux of
the monopoles is proportional to $12\pi$. Therefore, magnetic flux of the
monopoles is related to the difference between the mentioned above versions of
the Standard Model.

In order to illustrate the emergence of the additional $Z_3$ and $Z_2$
symmetries in the Standard Model we consider Petite Unification of strong and
Electroweak interactions discussed in \cite{PUT, PUT1}. In the mentioned papers
three possibilities to construct the unified theory at Tev were distinguished
among a number of various ones. Namely, let us consider the unified group to be
the product of $SU(4)_{PS}$ and $SU(N)^k$, where $SU(4)_{PS}$ unifies lepton
number with color as in Pati-Salam models \cite{PATI}. In the theory there are
two independent couplings $\alpha_s$ and $\alpha_W$ correspondent to the two
groups mentioned above. Then if we require that the spontaneous breakdown of
$SU(4)_{PS}\otimes SU(N)^k$ happens at a Tev scale we are left with the three
possibilities: ${\rm PUT}_0(N=2,k=4);\,{\rm PUT}_1 (N=2,k=3);\, {\rm
PUT}_2(N=3,k=2)$. The other choices of $N$ and $k$ cannot provide acceptable
values of coupling constants at the Electroweak scale.

It will be useful to represent the breakdown pattern correspondent to the
models $\rm PUT_0, PUT_1, PUT_2$ in terms of the loop variables $\Gamma, U$,
and $\theta$ calculated along the arbitrary closed contour $\cal C$.

In $\rm PUT_2$ at the Electroweak scale $SU(4)_{PS}\otimes SU(3)^2$ parallel
transporter $\Omega$ along the contour $\cal C$ is expressed through $\Gamma,
U$, and $\theta$ as follows:

\begin{equation}
\Omega = \left( \begin{array}{c c}

\Gamma^+ e^{\frac{2i\theta}{3}} & 0  \\
0 & e^{-2i\theta}

\end{array}\right) \otimes\left( \begin{array}{c c c}

e^{\frac{-4i\theta}{3}} & 0 & 0 \\
0 & e^{\frac{2i\theta}{3}} & 0 \\
0 & 0 & e^{\frac{2i\theta}{3}}

\end{array}\right)\otimes\left( \begin{array}{c  c}

U e^{-\frac{i\theta}{3}} & 0  \\
0 & e^{\frac{2i\theta}{3}}

\end{array}\right)\label{PUT2}
\end{equation}

From (\ref{PUT2}) it is  straightforward that values (\ref{Sing}) of the Wilson
loops $\Gamma$, $U$, and $e^{i\theta}$ with $N = 0, 3 \in Z_2$ lead to $\Omega
= {\bf 1}$. The field strength of the $SU(4)_{PS}\otimes SU(3)^2$ gauge field
is expressed through $\Omega$ calculated along the infinitely small contour.
Then the pure gauge field action in the low energy limit (at the Electroweak
scale) is invariant under an additional $Z_2$ symmetry. This means that in $\rm
PUT_2$ actual breakdown pattern is $SU(4)_{PS}\otimes SU(3)^2 \rightarrow
SU(3)\times SU(2) \times U(1)/Z_2$ and not $SU(4)_{PS}\otimes SU(3)^2
\rightarrow SU(3)\times SU(2) \times U(1)$. Therefore, we expect $Z_2$
monopoles to exist in this unified model with the masses of the order of $40*3
= 120$ Tev.

Here we used the values of Electroweak charges calculated in \cite{PUT} in
order to represent the breakdown pattern in a form useful for our purposes. One
can check directly that the gauge group element of the form (\ref{PUT2}) acts
appropriately on the Standard Model fermions arranged in the representations
listed in \cite{PUT}. The same check could be performed also for the models
$\rm PUT_1$ and $\rm PUT_0$ considered below.

In $\rm PUT_1$ at the Electroweak scale $SU(4)_{PS}\otimes SU(2)^3$ parallel
transporter $\Omega$ along the contour $\cal C$ is expressed as follows:

\begin{equation}
\Omega = \left( \begin{array}{c c}

\Gamma^+ e^{\frac{2i\theta}{3}} & 0  \\
0 & e^{-2i\theta}

\end{array}\right)\otimes U \otimes \left( \begin{array}{c  c}

e^{-i\theta} & 0  \\
0 & e^{i\theta}

\end{array}\right) \otimes\left( \begin{array}{c  c}

e^{i\theta} & 0  \\
0 & e^{-i\theta}

\end{array}\right)\label{PUT1}
\end{equation}

It is  straightforward that values (\ref{Sing}) of the Wilson loops $\Gamma$,
$U$, and $e^{i\theta}$ with $N = 0, 2, 4 \in Z_3$ lead to $\Omega = {\bf 1}$.
This means that in $\rm PUT_1$ actual breakdown pattern is $SU(4)_{PS}\otimes
SU(2)^3 \rightarrow SU(3)\times SU(2) \times U(1)/Z_3$ and not
$SU(4)_{PS}\otimes SU(2)^3 \rightarrow SU(3)\times SU(2) \times U(1)$. Thus,
$Z_3$ monopoles should exist in this unified model with the masses of the order
of $40*2 = 80$ Tev.

In $\rm PUT_0$ at the Electroweak scale $SU(4)_{PS}\otimes SU(2)^4$ parallel
transporter $\Omega$ along the contour $\cal C$ is expressed through $\Gamma,
U$, and $\theta$ as follows:

\begin{equation}
\Omega = \left( \begin{array}{c c}

\Gamma e^{\frac{i\theta}{3}} & 0  \\
0 & e^{-i\theta}

\end{array}\right)\otimes U \otimes\left( \begin{array}{c  c}

e^{i\theta} & 0  \\
0 & e^{-i\theta}

\end{array}\right)\otimes U \otimes\left( \begin{array}{c  c}

e^{i\theta} & 0  \\
0 & e^{-i\theta}

\end{array}\right)\label{PUT0}
\end{equation}

One can easily find that in $\rm PUT_0$ actual breakdown pattern is
$SU(4)_{PS}\otimes SU(2)^4 \rightarrow SU(3)\times SU(2) \times U(1)/Z_3$ and
not $SU(4)_{PS}\otimes SU(2)^4 \rightarrow SU(3)\times SU(2) \times U(1)$.
Thus, in this model $Z_3$ monopoles also should exist with the masses of the
order of $80$ Tev. It should be mentioned here that $\rm PUT_0$ seems to be
excluded due to the extremely high value of branching ratio for the process
$K_L \rightarrow \mu e$.

Thus, we considered the Standard Model embedded into a unified model, the
symmetry of which breaks down to the gauge group of the SM at a few Tev. During
the breakdown monopoles may appear, which have masses of the order of $40$ Tev.
Those objects could become the lightest topologically stable magnetic
monopoles. In principle, during the high energy collisions monopole -
antimonopole pair may appear. The size of the monopoles should be of the order
of $1$ $\rm Tev^{-1}$ while the typical size of the objects to be created
during the collision should be about  $80$ $\rm Tev^{-1}$. Therefore, the
appearance of monopole - antimonopole pair should be suppressed by a function
of their ratio. The complete theoretical consideration of the monopole -
antimonopole pair creation is a still unresolved problem even for the usual
t'Hooft - Polyakov monopooles (see, for example, \cite{MONOPOLE} and references
therein). Therefore, we cannot give in the present paper any estimate of the
cross sections  correspondent to the creation of  monopoles in high energy
collisions.

The appearance of the mentioned monopoles in the early Universe may have
certain cosmological consequences, observation of which could become the way to
distinguish between the theories with the gauge groups $SU(3)\times SU(2)
\times U(1)$ and $SU(3)\times SU(2) \times U(1)/{\cal Z}$. In particular, we
expect, that the spontaneously  broken symmetry of the unified model is
restored at high temperature. We expect that in the early Universe at the
temperatures close to the temperature $T_c$ of the correspondent transition
 monopoles considered in this paper could appear in the elementary
processes with a high probability. Those monopoles may even be condensed at
$T>T_c$ as Nambu monopoles at the temperatures above the Electroweak transition
temperature \cite{BVZ2006}. Then,  monopoles may play an important role, say,
in the processes with changing of baryon number \cite{Rubakov}.

This work was partly supported  by RFBR grants 06-02-16309, 05-02-16306, and
07-02-00237.

\clearpage

\end{document}